\documentclass[12pt]{iopart}

\usepackage{graphicx}
\usepackage{dcolumn}
\usepackage{bm}
\usepackage{color}
\usepackage[utf8x]{inputenc}
\usepackage[pagewise]{lineno}

\usepackage[bookmarks=false, urlcolor=blue]{hyperref}

\begin{document}

\title[Deep Learning-Assisted Classification of Quantum Gas Microscope Images]{Deep Learning-Assisted Classification of Site-Resolved Quantum Gas Microscope Images}

\author{Lewis R. B. Picard$^1$, Manfred J. Mark$^{2,3}$, Francesca Ferlaino$^{2,3}$, Rick van Bijnen$^{2,4}$}
\address{$^1$
 Department of Physics, Harvard University, Cambridge, Massachusetts, 02138, USA.
}
\address{$^2$
Institut f{\"u}r  Quantenoptik  und  Quanteninformation, {\"O}sterreichische  Akademie  der  Wissenschaften,  6020  Innsbruck, Austria}
\address{$^3$
Institut f{\"u}r  Experimentalphysik und  Zentrum  f{\"u}r Quantenoptik, Universit{\"a}t  Innsbruck,  6020  Innsbruck, Austria}
\address{$^4$ Center for Quantum Physics, University of Innsbruck, Austria}
\ead{lewispicard@g.harvard.edu}

\date{\today}

\begin{abstract}

We present a novel method for the analysis of quantum gas microscope images, which uses deep learning to improve the fidelity with which lattice sites can be classified as occupied or unoccupied. Our method is especially suited to addressing the case of imaging without continuous cooling, in which the accuracy of existing threshold-based reconstruction methods is limited by atom motion and low photon counts. We devise two neural network architectures which are both able to improve upon the fidelity of threshold-based methods, following training on large data sets of simulated images. We evaluate these methods on simulations of a free-space erbium quantum gas microscope, and a noncooled ytterbium microscope in which atoms are pinned in a deep lattice during imaging. In some conditions we see reductions of up to a factor of two in the reconstruction error rate, representing a significant step forward in our efforts to implement high fidelity noncooled site-resolved imaging.
\end{abstract}

\\
\\
\textit{Published as}: Picard, L. R. B., Mark, M. J., Ferlaino, F. and van Bijnen, R. Deep learning-assisted classification of site-resolved quantum gas microscope images.  \textit{Meas. Sci. Technol.} \textbf{31} 025201, 2020. 
\\
DOI: \href{https://doi.org/10.1088\%2F1361-6501\%2Fab44d8}{10.1088/1361-6501/ab44d8}.

\maketitle
\ioptwocol

\section{\label{sec:Intro} Introduction}

Over the past decade, site-resolved fluorescence imaging of atoms in optical lattices has become an essential tool for researchers working in ultracold atomic physics and quantum simulation \cite{gross_quantum_2017}. The adoption of this powerful technique has been driven by improvements in both high-resolution imaging systems and computational techniques for identifying atoms separated by distances close to or below the diffraction-limited resolution \cite{karski_nearest-neighbor_2009, sherson_single-atom-resolved_2010}. The task of site-resolved imaging consists of two distinct parts: 1) building an imaging system which is able to detect multiple fluorescence photons scattered by each atom in an optical lattice and 2) analyzing the recorded image in order to determine whether or not each lattice site is occupied by an atom. This is both an experimental challenge, constructing a high resolution microscope, and a computational one, devising an algorithm to reliably reconstruct the underlying lattice occupation from the recorded image. At present, the range of species that can be imaged remains limited by the need to continuously cool atoms during fluorescence imaging. In the vast majority of existing site-resolved imaging experiments, atoms are pinned in place by a deep lattice and continuously laser-cooled during imaging \cite{sherson_single-atom-resolved_2010,bakr_quantum_2009,cheuk_quantum-gas_2015,edge_imaging_2015,haller_single-atom_2015,omran_microscopic_2015,parsons_site-resolved_2015, mitra_quantum_2018}. In this case the distribution of bright pixels in a fluorescence image ideally results only from the point-spread function (PSF) of the imaging system. Imaging without cooling limits the number of photons which can be detected from each atom, which gets rapidly heated up and displaced from its original position by scattering of the imaging light. This heating reduces the fidelity of traditional threshold-based reconstruction methods. Here, we propose a novel method of analyzing fluorescence images of atoms in optical lattices using deep learning, in order to improve the performance of imaging without continuous cooling.

\begin{figure}
    \centering
    \includegraphics[width=0.48\textwidth]{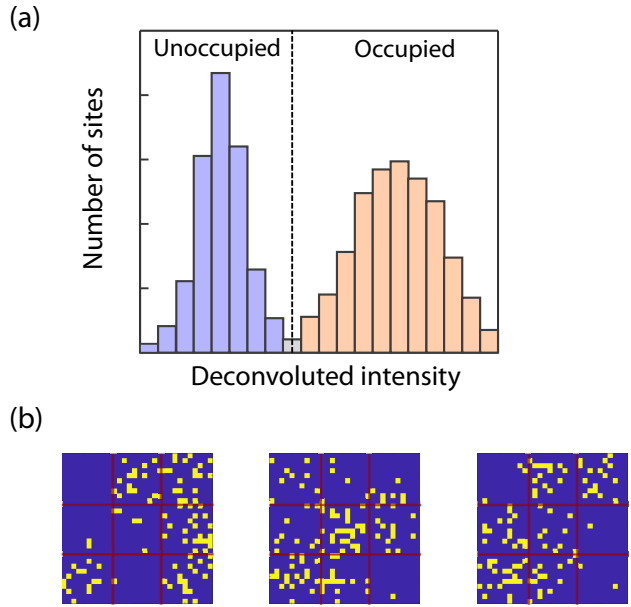}
    \caption{(a) Illustration of threshold-based reconstruction. A histogram of intensities following deconvolution at each site in a set of images is plotted. If sites are separated by more than or close to the diffraction-limited resolution, this will reveal a bimodal distribution of intensities. The threshold intensity used to classify a site is determined by the point at which the two peaks overlap. (b) Examples of simulated images of three-by-three erbium lattice segments, with a lattice constant of 266 nm and 1.5 $\mu$s illumination time. The superimposed red lines indicate the lattice site boundaries. Of the three images, only the center one has an occupied central lattice site.} 
    \label{fig:naive_hist}
\end{figure}

 The most widely used method for reconstruction of the lattice occupation pattern in existing experiments requires first deconvoluting each image with the known PSF of the imaging system. This PSF can be determined experimentally by averaging raw images of many isolated atoms, or calculated based on known optical parameters of the imaging system \cite{karski_nearest-neighbor_2009, sherson_single-atom-resolved_2010,bakr_quantum_2009,cheuk_quantum-gas_2015,edge_imaging_2015,haller_single-atom_2015,omran_microscopic_2015,parsons_site-resolved_2015}. Deconvolution allows a single value of the light intensity to be determined for each lattice site. The distribution of light intensities will generally consist of two distinct peaks corresponding to occupied and unoccupied sites, as illustrated in figure \ref{fig:naive_hist}(a). The degree of overlap of the histogram peaks is determined both by the background noise level and the overlap of point-spread functions of atoms on neighbouring sites. The bimodal distribution is eventually washed out entirely for high noise levels and/or for atom separations significantly below the width of the point spread function of the imaging system.  Taking a large enough sample of lattice sites allows the estimation of the underlying distribution, from which a single threshold value can be derived which can be used to classify the occupation of all sites \cite{bakr_quantum_2009,cheuk_quantum-gas_2015,edge_imaging_2015,nelson_imaging_2007}. Some variations on this basic method exist, such as determining the occupation by minimizing the difference between a real image and a reconstruction generated through convolution with the PSF \cite{sherson_single-atom-resolved_2010}, but the experimental requirements remain similar. More recent work on parametric deconvolution, described in \cite{alberti_super-resolution_2016}, has shown that a more sophisticated model which uses knowledge of both the point spread function and the restricted geometry of the lattice can improve the discrimination of nearby atoms.

Without continuous cooling, atoms will be significantly heated during the imaging process. This heating occurs through the build-up of velocity kicks an atom receives each time it absorbs and re-emits a photon, eventually giving it enough kinetic energy to escape the potential well of a lattice site. Cooling and confinement by a deep pinning lattice allows the capture of images consisting of hundreds of scattered photons per atom, with reconstruction fidelity limited mainly by atom losses and hopping between lattice sites \cite{greif_site-resolved_2016}. Implementing continuous cooling is, however, among the more experimentally challenging facets of a single-site imaging system. The requirement of a cooling transition which can simultaneously be used for imaging severely limits the range of species which can be imaged, and generally requires that a quantum gas microscope is custom-built for each new species. As a result, the extension of single-site imaging to fermionic alkaline atoms came significantly later than boson-imaging, requiring the implementation of more sophisticated cooling techniques, such as Raman sideband and EIT cooling \cite{gross_quantum_2017,cheuk_quantum-gas_2015,haller_single-atom_2015}. These cooling techniques tend to increase experimental complexity, needing additional cooling beams, and, in the case of EIT cooling, may themselves introduce high levels of background light, which must then be reduced by other means, such as alternating cooling and imaging pulses in a single imaging cycle \cite{edge_imaging_2015}. To our knowledge only one example of optical lattice imaging without cooling has been published at this time, which relies on confining Yb atoms in a deep lattice and using short imaging pulses to prevent losses due to heating \cite{miranda_site-resolved_2017}. Fluorescence imaging of single Li atoms in free flight has recently been achieved, but using this method multiple atoms can only be reliably resolved at a separation greater than 32 $\mu$m, precluding the study of short-scale many-body dynamics \cite{bergschneider_spin-resolved_2018}.

We propose a method for reconstructing optical lattice images to single-site resolution which does not require atoms to be confined to a lattice site during imaging. When atoms are neither continuously cooled nor pinned by a deep lattice, they will move away from their original lattice site on a random walk as they scatter photons from the imaging beam. High-resolution imaging without extra cooling and optical pinning will bring enormous experimental and conceptual simplification, and will be essential to the development of ultrafast microscopy. In this respect, atoms with strong optical transitions for imaging and large masses, such as lanthanides, are perfect candidates, and are a target of growing interest as many-body quantum systems in the community. In the case of our planned Er microscope, the lattice will be switched off entirely during imaging, allowing the atoms to diffuse in free space. In other cases, such as the Yb lattice experiment that we simulate to assess our networks, the lattice potential is deepened during imaging to provide some confinement without cooling, such that atoms jump between lattice sites as they heat up \cite{miranda_site-resolved_2017}.

The random motion of the atoms makes the reconstruction of the lattice occupation an intractable inverse problem, meaning that there is no way to exactly determine the most likely initial atom distribution which gave rise to a particular recorded image. It is nevertheless possible to approximate the atom as a fixed point emitter, with an effective PSF broadened by atom motion compared to the true optical PSF. This method may be sufficient when lattice spacings are large compared to the atom displacements, or when many photons are collected before the atoms move away from their starting positions. However, an additional restriction imposed by noncooled imaging is that the total photon count must be small, as only a few photons can be detected before atoms move too far to be distinguished from their neighbours, severely limiting the applicability of the stationary emitter approximation. We suggest that deep neural networks provide a way to overcome some of the limitations of noncooled image reconstruction. The advantage of using deep learning for data-analysis lies in the fact that a deep neural network can approximate non-linear relationships between input data. This is especially useful in the analysis of intractable inverse problems. In the past few years, machine learning has found an increasing number of applications in physics, particularly in classification problems \cite{zdeborova_machine_2017}. Deep neural networks may offer advantages in both speed and accuracy over existing approximations, as has been demonstrated for a range of physical problems, including determining observable properties of electrons in arbitrary 2D potentials \cite{mills_deep_2017}, reading out trapped ion qubits \cite{seif_machine_2018} and reconstructing the optical phase of imaging light at an objective from low photon count recorded images \cite{goy_low_2018}. In other cases they may allow classification of experimental data for which no agreed-upon approximate model exists, which has led to their use in identifying phase transitions in quantum many-body systems \cite{chng_machine_2017,carrasquilla_machine_2017, rem_identifying_2019} and evaluating theoretical models of interactions of fermions in an optical lattice \cite{bohrdt_classifying_2019}. Outside the realm of classification problems, recent work has focused on the rich field of unsupervised machine learning, in which models are trained with unlabelled data based on some metric internal to the data set, such as the degree to which different inputs can be divided into non-overlapping clusters \cite{carleo_machine_2019}. Unsupervised learning has recently been demonstrated to be useful in quantum state tomography, where neural network states representing the amplitude and phase of a many-body quantum system are learned based on sets of measurements of its state in a range of bases \cite{torlai_neural-network_2018}.

The reconstruction procedure we describe here has been designed primarily to analyze images from our planned noncooled erbium quantum gas microscope \cite{ilzhofer_two-species_2018}, but is generally applicable to most cooled and noncooled imaging systems. To illustrate the task at hand, figure \ref{fig:naive_hist}(b) shows some typical (simulated) example images that our method aims to classify. In the present paper we test two different deep learning classifiers of different levels of complexity, and compare their performance to a threshold-based reconstruction model.

\section{\label{sec:NNet}Reconstruction Using Deep Learning}

\begin{figure}
    \centering
    \includegraphics[width=0.48\textwidth]{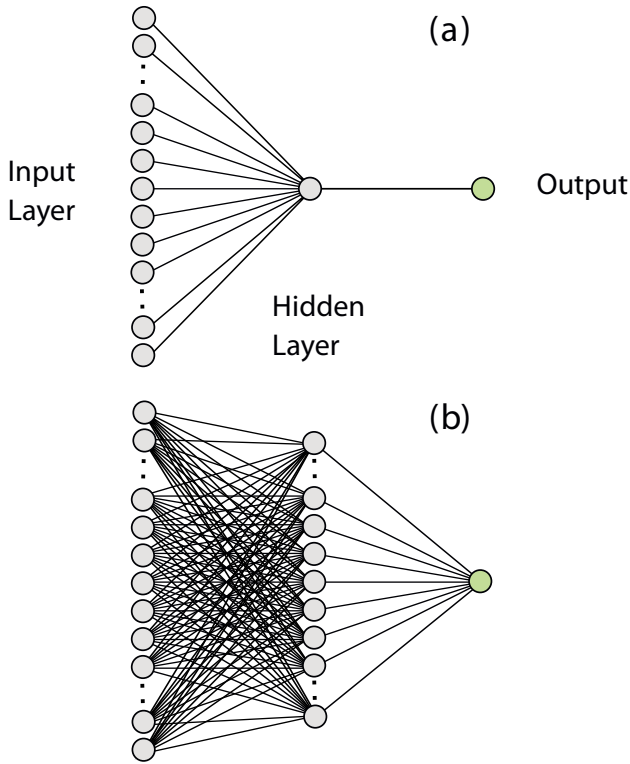}
    \caption{Illustration of three-layer feedforward neural network architectures with all-to-all interlayer connections in $N$-1-1 (a) and $N$-$M$-1 (b) configurations, where $N$ is the number of pixels in an input image and $M$ is the number of neurons in a hidden layer.} 
    \label{fig:nets_illustr}
\end{figure}

Deep neural networks are generally models that transform an input vector, in our case an array of pixels, into an output vector. In this case the output is a scalar value indicating whether or not a lattice site is occupied. Deep neural networks perform their function using a series of two or more consecutive transformations, each of which takes the output of the previous one as its input \cite{goodfellow_deep_2016}. The transformations are said to connect different layers of the network, beginning with the input layer, consisting of a raw input vector, through to the final output layer. A hidden layer is one which lies between the input and the output, whose state is not read out to the user. The model as a whole is referred to as an artificial neural network, as its structure is inspired by, though not actually very similar to, biological neural networks \cite{bengio_towards_2015}. Each element of a layer, usually a scalar number, can be referred to as a neuron. The parameters of the network that define the precise mapping from one layer to the next can be \textit{learned} by repeatedly evaluating the performance of the network on a set of test input vectors, and adjusting the parameters accordingly.

A feedforward neural network, illustrated in figure \ref{fig:nets_illustr}, is among the simplest neural network architectures that exist. It consists of a series of layers, where each neuron in a layer is connected to every neuron of its neighbouring layers, and there are no intralayer connections. The action of the network on an input data vector is, in its most basic form, a series of matrix multiplications. Generally a bias vector is also added to the output of each layer, and a transfer function may also be applied to each output. Thus, the action of a single layer can be written as
\begin{equation}
    \label{feedforward_layer}
    \mathbf{y}^{(i)} = f(W^{(i)}\mathbf{y}^{(i-1)} + \mathbf{b}^{(i)})
\end{equation}
where $\mathbf{y}^{(i-1)}$ is an $m$ element input vector representing the neuron values of layer $i-1$, $\mathbf{y}^{(i)}$ is an $n$ element output representing the neuron values of layer $i$, $W^{(i)}$ is an $n\times m$ matrix, $\mathbf{b}^{(i)}$ is an $n$ element bias vector and $f$ is an arbitrary transfer function applied element-wise to the the intermediate value to give an output neuron state. The transfer function is often used to map scalar values back to the interval $\{0, 1\}$. The process of training a neural network broadly consists of adjusting weight matrices and bias vectors to optimize the output for a particular problem.  The performance of a trained network can then be evaluated by measuring its generalization error, the rate at which it correctly classifies items in a previously unseen data set. In principle, a two-layer feedforward network is capable of learning any arbitrary relationship between elements of an input data vector \cite{goodfellow_deep_2016}.  In practice it is often difficult to train such a network, particularly when dealing with large input vectors, such as the high-magnification images of lattice segments we use to train our classifier.

Below, we discuss a number of neural network architectures with which we have experimented in order to classify lattice images. All of our neural networks are trained on large data sets of simulated images of three-by-three lattice site regions (see Appendix for discussion of the simulation). The reason for using three-by-three segments is that these are able to capture the first-order correlations between the brightness of a lattice site and its eight nearest neighbours while still being small enough that we can simulate training data sets in which every possible arrangement of atoms is represented. When the networks are applied to test images, these are first broken down into overlapping three-by-three segments, which are then individually fed into the network for classification of the central site of each segment.

\subsection{\label{sec:trilayer}Threshold reconstruction as a three-layer network}


In order to better understand the process of neural network training and how it can be used to achieve improvements in fidelity, we first wish to trace a direct link between threshold reconstruction and some simple neural network architectures for which we can provide qualitative \textit{post-hoc} interpretations \cite{lipton_mythos_2016}. To this end, we implement a basic form of threshold reconstruction in a form resembling a neural network, and compare it to an equivalent neural network trained on a data set of simulated images.

\begin{figure}
    \centering
    \includegraphics[width=0.48\textwidth]{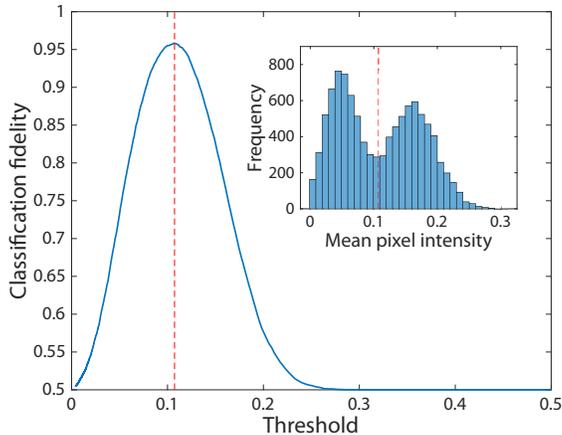}
    \caption{Range of fidelities achievable using weighted sum threshold-based reconstruction by varying the mean pixel intensity threshold, which is equivalent to $b/w$ in the neural network representation. Fidelities are evaluated on a data set of simulated images of unconfined erbium atoms in a lattice of period 266 nm, generated according to the procedure described Appendix A. The inset shows the bimodal distribution of mean pixel intensities in the simulated data set. Prior to optimizing the threshold the pixels are weighted by a PSF centered on the lattice site, which improves the separation of the peaks in the intensity distribution. In both plots, the optimal threshold of 0.108 is indicated by a dashed red line.}
    \label{fig:naive_matrix}
\end{figure}

The simplest way to determine an intensity value for threshold reconstruction is to simply add up all the bright pixels in a lattice site. This could be trivially represented in the feedforward neural network form given in equation 1 through multiplication of the input by an $m\times 1$ binary vector, with a one multiplying every pixel in the region to be summed and zeros everywhere else: $ \mathbf{y}^{(i)} = \mathbf{w}\cdot\mathbf{y}^{(i-1)}$. To improve the discrimination between photons from lattice atoms and noise counts, one can replace the simple sum by a weighted sum using a PSF centered on the lattice site being classified. The lattice spacing and alignment can be determined experimentally beforehand by various means, such as Fourier transforming a whole lattice image \cite{omran_microscopic_2015} or projecting images onto each axis of the imaging plane and fitting with a periodic series of Gaussians \cite{karski_nearest-neighbor_2009, sherson_single-atom-resolved_2010}, as we do in this work. The sum of pixels weighted by the PSF can then be expressed as a row-matrix multiplication linking the input layer and single-neuron hidden layer of a neural network. In other words, the matrix $W^{(i)}$, for $i = 1$, in equation (\ref{feedforward_layer}) is simply a row vector $W^{(1)}_{1j} = \mathrm{PSF}(\mathbf{x}_j)$, with $\mathbf{x}_j$ the coordinates of the $j$-th pixel of the PSF. The transfer function applied at the hidden layer is $f(x) = x$. The transformation from the hidden layer to the single-neuron output consists of a scalar multiplication by a weight $w$ followed by the addition of a bias $b$ and application of the logistic-sigmoid transfer function $f(y) = (1 + \exp(-y))^{-1}$, producing an output in the range 0 to 1, with 0 corresponding to an unoccupied central site and 1 corresponding to occupied. This layer performs the same role as the comparison of site intensity to a fixed threshold. In principle the above network could also be reduced to a two-layer network, but for later convenience we employed a three-layer format. By scanning the parameter $w$, the maximum possible fidelity of the weighted sum threshold reconstruction can be determined, as illustrated for the case of noncooled erbium atoms in figure \ref{fig:naive_matrix}.

\begin{figure}
    \centering
    \includegraphics[width=0.48\textwidth]{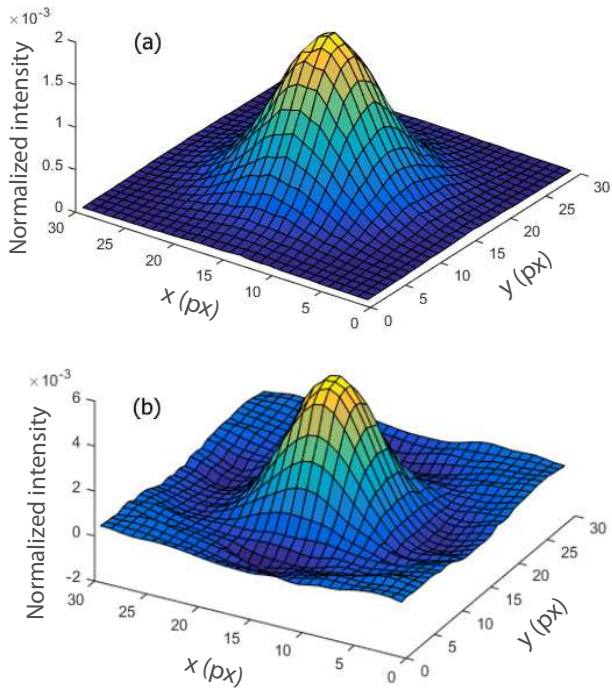}
    \caption{(a) Point spread function determined by averaging simulated images of 10000 isolated Er atoms, with a 3 $\mu$s imaging pulse, used to weight pixels in an input image for threshold-based reconstruction. (b) Learned pixel weights after training the three-layer network in figure \ref{fig:nets_illustr}(a) on a data set of 102400 distinct images, using the PSF as the initial state of the input layer. Without any additional human input, the network learns to assign a negative weight to bright pixels in the neighbouring sites of the central atom. This example illustrates the ability of even very simple neural networks to learn approximate models of the correlations between neighbouring lattice sites.}
    \label{fig:compare_visible}
\end{figure}

We can gain some insight into the neural network training process by training a network using the optimal weighted sum threshold as our initialization condition. As a first step, we leave the network architecture fixed, but optimize the weight matrix $W^{(1)}$ and weight $w$ using conjugate gradient descent, training the network with a set of simulated images, after initializing $W^{(1)}$ with the PSF reshaped to a row vector as described above. During the training process, the weights assigned to pixels in the input image and the classification threshold are adjusted so as to minimize the reconstruction error. Given that the hidden layer has only a single neuron, this still directly corresponds to a weighted sum of all the pixels in an input image. What we see, however, is that during the training procedure the neural network learns to negatively weight bright pixels in neighbouring lattice sites, and achieves a significant increase in fidelity as a result, as illustrated in figure \ref{fig:compare_visible}. This tells us that without additional manual intervention on the part of researchers the network can learn to compensate for overlap of signals from filled lattice sites onto their neighbours. We also found that while manually initializing the network with the PSF allows it to reliably converge to a good classifier, using any random initialization generally does not converge to a good solution. This shows that though training even this simple neural network leads to an improvement over the manually optimized method, it remains very sensitive to user defined initialization conditions, which are specific to each imaging system.

The weighted sum model alone does not represent the best available form of threshold reconstruction. Deconvolution, or equivalently fitting an image with a set of Gaussians centered on each lattice site, is the most widely used method, described in \cite{ parsons_site-resolved_2015,greif_site-resolved_2016, miranda_site-resolved_2017}, among others. A threshold can then be applied to the fit amplitude of each Gaussian to assign the sites as occupied or unoccupied. The fit with a joint distribution of multiple Gaussians serves the purpose of discriminating between the signals produced by atoms on neighbouring lattice sites. An increased amplitude for a Gaussian centered on one site generally corresponds to a reduction of the amplitude on its neighbours, representing a reduced occupation probability. Ideally, this method converges to the most likely distribution of lattice site occupations for the whole image. We implement this method for three-by-three lattice segments as the state-of-the-art benchmark against which we compare our machine learning methods. For the tight atom confinement and larger lattice spacings ($\geq512$ nm) typical of existing quantum gas microscopes, threshold reconstruction remains highly effective \cite{sherson_single-atom-resolved_2010,bakr_quantum_2009,cheuk_quantum-gas_2015}. We explore this regime by simulating imaging of erbium atoms in a two-dimensional square lattice with a spacing of 532 nm, under which conditions we see up to 99.9\% threshold-based reconstruction fidelity. Threshold fidelity drops off as the lattice spacing is decreased and PSF overlap increases, however, and is closer to 97\% in the 266 nm spacing system we aim to image.

In order to re-express the Gaussian fit method as a three-layer feedforward network, we use a 512-neuron hidden layer. Each of the neurons in the hidden layer is connected to the input image in the same way as the weighted sum  method described above, but now the weight matrices correspond not just to a single Gaussian PSF on the central site, but to sums of Gaussians on each site in all of the 512 possible distributions of occupied and unoccupied lattice sites. The distributions with the greatest overlaps with the real image will then produce greater activations in the corresponding hidden neurons. The initial weights to the final layer are then a sum of all the hidden neuron values corresponding to an occupied central site, minus all those corresponding to an empty central site. This is effectively a majority vote among all the possible Gaussian fits as to whether the central site is occupied. The output is then normalized to provide a value in the range \{0,1\}.  This architecture is illustrated in figure \ref{fig:nets_illustr}(b). As always, the network is then trained to optimize fidelity from these initial conditions. In section \ref{sec:performance} we refer to this network architecture by the name ``Gaussian network" when we compare it to both the manual Gaussian fit and the more sophisticated deep convolutional network described below.

We find that the output of the feedforward neural network is itself a good estimator of the confidence of the result. For example, an output of 0.6 has an approximately 60\% probability of genuinely corresponding to an occupied site. An output of 0.1 has a 90\% chance of corresponding to an unoccupied site. This allows the classifier to be easily used for confidence-weighting or post-selection of experimental results.

\subsection{\label{sec:autoenc}Convolutional neural network reconstruction}

The three-layer networks introduced above are based on the assumption that atoms act like point sources fixed at a lattice site. Without continuous cooling, however, atoms wander between sites, so ideally a model would be able to encompass the movement of atoms and distinguish between an atom originating at a central site and one which has been displaced there.
In order to produce a model that takes into account more than simply how well a central site is fit by a static PSF, we turn to a more sophisticated network architecture known as a convolutional neural network.

A convolutional neural network works on basis of convoluting an input image with a learned kernel, such that each neuron in a subsequent hidden layer corresponds to the convolution for a specific position of the kernel on the input. Rather than learning a single weight for every input pixel, as in a feedforward network, during training the convolutional network learns the weights of the kernel, which are then re-used for all the different subsections of the input. This is useful for identifying significant features which may occur at any position within a 2D image, such as the PSF of a freely wandering atom. In a realistic convolutional neural network architecture, multiple kernels are often used to identify different sets of features.

A deep convolutional neural network layers this process several times, learning one set of kernels for the input image, then another set with which to convolute the outputs of the first, etc. While the first kernels tend to represent visibly recognizable features in the input, the subsequent layers are more abstracted, learning, for example, to identify correlations between different features identified by the first layer. A convolution operation is usually accompanied by normalization and application of a function such as a rectified linear unit, serving much the same function as the transfer functions in feedforward neural networks. Most deep convolutional networks also include pooling layers, which perform the function of producing a statistical summary of the outputs of a convolutional layer. Common pooling processes include taking the maximum or the average value of the convolution outputs in a given region. Pooling can also perform the function of dimensionality reduction; if the overlap of pooling regions is reduced, the number of output neurons will be smaller than the number of inputs. This reduces the complexity of the next convolutional stage, increasing training speed and reducing memory usage. The number of pixels between each step of the pooling filter is known as the \textit{stride}. Finally, a convolutional network will usually finish with a fully connected layer, such as those illustrated in figure \ref{fig:nets_illustr}, which produces an output of a fixed size for a classification or regression task.

\begin{figure*}
    \centering
    \includegraphics[width=0.9\textwidth]{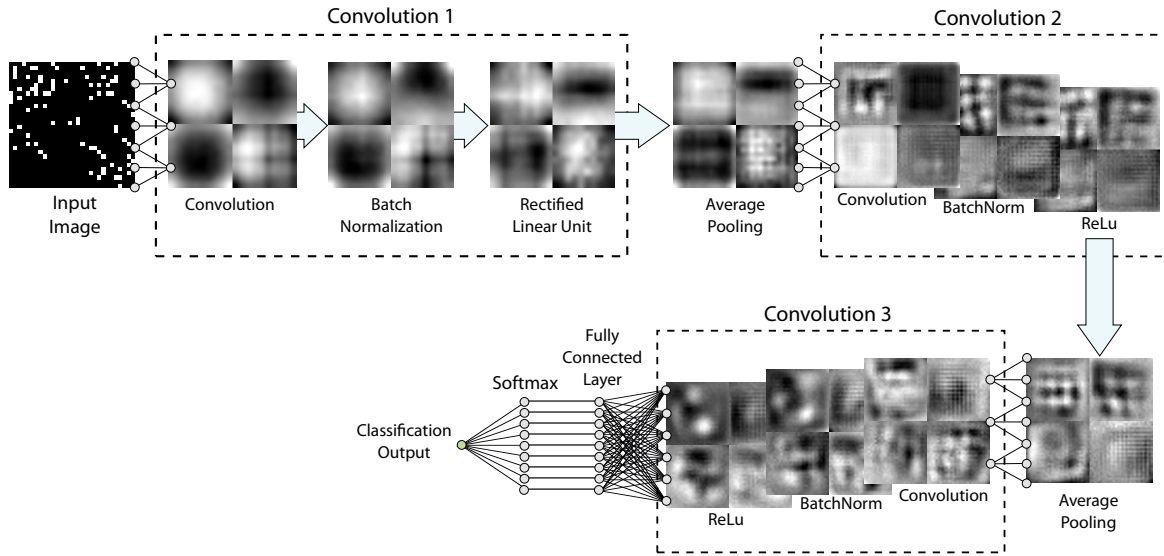}
    \caption{Illustration of the convolutional neural network architecture used in the present work. The images are a sample of the features learned at each layer of the network. These are created using a version of the deepDream algorithm in MATLAB \cite{matlab_deepdreamimage_2019}, shown as a grid of artificial images which most strongly activate those features.}
    \label{fig:conv_illustr}
\end{figure*}

In this work, we use a network with three convolutional layers and two average pooling layers, followed by a fully connected classification layer. This network architecture is illustrated in figure \ref{fig:conv_illustr}, along with visual representations of the features learned by the network trained on simulated lattice images. By testing a range of network parameters, we find we can achieve optimal performance with a convolution kernel size of 10-by-10 pixels, corresponding to the size of a single lattice site in our training images. We also optimize the size and stride of pooling layers, and the number of training images, all of which is detailed in Appendix B.

\section{\label{sec:performance}Evaluating Classifier Performance}

We evaluate the performance of both the Gaussian and convolutional networks introduced in the previous section in a range of different simulated experimental conditions, and compare them against the benchmark of Gaussian fit amplitude threshold-based reconstruction. The neural network classifiers are extremely flexible, and can be applied to the analysis of any two-dimensional lattice images, provided the imaging system is understood well enough to simulate the imaging of three-by-three lattice segments to generate labelled training data. Further details of our simulation of the imaging system are provided in the Appendix. The performance metric we use is the reconstruction fidelity across the whole lattice, i.e., the percentage of sites which are correctly classified when the reconstruction method is used to assign every site in a previously unseen lattice image. Depending on the particular experimental context in which these methods are applied, other performance metrics could be more appropriate. In investigations of Mott-insulating behaviour, for example, the rate at which a classifier correctly identifies holes in an otherwise uniformly filled lattice could be a more useful metric \cite{greif_site-resolved_2016, bohrdt_classifying_2019}.

\subsection{\label{sec:erbium}Noncooled erbium lattice}

\begin{figure}
    \centering
    \includegraphics[width=0.48\textwidth]{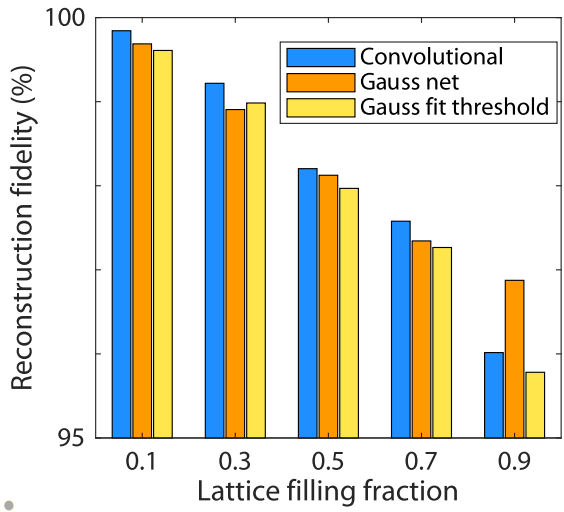}
    \caption{Fidelities of three reconstruction methods, for various lattice filling fractions. From left to right, the methods are: convolutional network, three-layer Gaussian network and threshold reconstruction. All test images are of unconfined erbium atoms at 266 nm spacing and 1.5 $\mu$s illumination time. For each filling fraction we simulate a ten-by-ten site lattice of a given filling, which we break up into overlapping three-by-three segments for fitting.}
    \label{fig:fidel_filling}
\end{figure}

We first test our model on the challenging case of noncooled and unpinned ultracold atoms. As a species of interest we choose erbium, which is a highly magnetic lanthanide atom that has recently been brought to quantum degeneracy \cite{aikawa_bose-einstein_2012,aikawa_reaching_2014}. We simulate the following experimental conditions: prior to imaging, Er atoms are held in a three-dimensional optical lattice with typical spacing of 266 nm. The lattice is then switched off and atoms are illuminated with a resonant light pulse of 1.5 $\mu$s. The atomic fluorescence is projected onto a CCD camera by our imaging system with a numerical aperture (NA) of 0.89. The imaging light operates on the 401 nm transition, for which we predict a maximum scattering rate, limited by the transition's natural linewidth, of $9.5\times10^7$ $\mathrm{ s}^{-1}$. With an imaging beam intensity $\sim$10 times higher than the saturation intensity of the transition, we expect to collect less than 90 photons per atom in a single image. Given this relatively small number of collected photons, we can reliably assume that a negligible number of pixels will be multiply illuminated, allowing us to binarize our images, facilitating the convergence of neural network training. For cases where the magnification is small enough compared to the lattice spacing that multiple illumination of pixels is likely, we have devised an alternative normalization function, given in equation A.1 in the appendix, to map the input to the range \{0,1\}.

We test convolutional network, Gaussian network and threshold reconstructions on previously unseen simulated images of entire lattices, which are divided into overlapping three-by-three site blocks for input to the networks. In figure \ref{fig:fidel_filling}, the fidelities of the various methods for a range of site occupation densities at 266 nm spacing, from a sparsely filled to an almost completely filled lattice, are shown. In the maximum uncertainty case of half filling, the error rate is reduced from 2.03\% for the Gaussian fit threshold method to 1.80\% for the convolutional network. For sparse filling the error rate of the convolutional network is just 0.16\%, while that of the threshold method is 0.39\%, a more than twofold improvement.  As the filling increases the performance of all methods decreases as a result of the  reduced distinguishability of individual occupied and unoccupied sites, even as the overall entropy of the entire lattice configuration decreases. We note that at high filling the 512 hidden neuron Gaussian network performs particularly well, better than both the convolutional network and the threshold-reconstruction, though we have no clear interpretation for this boost in performance.

As we increase the lattice period, reconstruction performance increases rapidly. The convolutional network achieves as high as 99.90\% reconstruction fidelity at a spacing of 532 nm and half-filling of the lattice. Threshold-based reconstruction in these conditions provides average fidelity of 99.83\%, indicating that the neural network continues to provide a small but significant advantage at high spacing. The convolutional network fidelity of 99.9\% is maintained at 0.1 lattice filling fraction, dropping only slightly to 99.5\% at 0.9 filling. This fidelity is achieved despite expected atom losses of $\sim3\%$ during imaging caused by atoms escaping the not fully closed imaging transition cycle. That is, the network is able to reliably identify most lost atoms even from the small number of photons they scatter prior to loss. 

\begin{figure}
    \centering
    \includegraphics[width=0.48\textwidth]{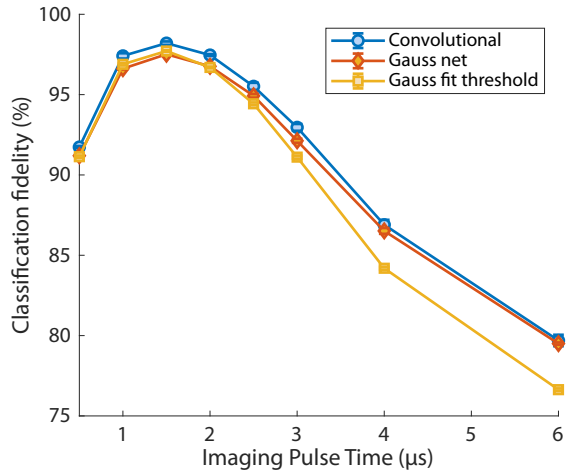}
    \caption{Reconstruction fidelities using convolutional network, three-layer Gaussian network and Gaussian fit threshold reconstruction for noncooled erbium lattices at a range of imaging pulse lengths, with a 266 nm lattice spacing and half-filling. The greatest fidelity is expected for a 1.5 $\mu$s imaging pulse.}
    \label{fig:imaging_pulse}
\end{figure}

We also use our simulation to estimate the imaging pulse time which maximizes fidelity. Figure \ref{fig:imaging_pulse} shows how the fidelity of threshold, three-layer and convolutional reconstruction changes with imaging pulse time. Simulations suggest that the highest reconstruction fidelity can be achieved for a 1.5 $\mu$s imaging pulse. It is assumed that at this timescale background light is not a significant contributor to image noise, so the noise level is taken to be constant over all pulse lengths. We observe that the performance of the threshold-based reconstruction drops off more rapidly with increased imaging time than the neural network methods, while the performances of the Gaussian and convolutional networks appear to converge. The simulations in figure \ref{fig:imaging_pulse} are conducted at half-filling of the lattice. We find that the fidelity drops off more sharply as imaging time is increased for dense filling of the lattice, though it plateaus at the filling percentage for greater than half-filling, corresponding to the error rate incurred by assigning all sites as occupied.

\subsection{\label{sec:Kozuma}Noncooled ytterbium in pinning lattice}

\begin{figure}
    \centering
    \includegraphics[width=0.48\textwidth]{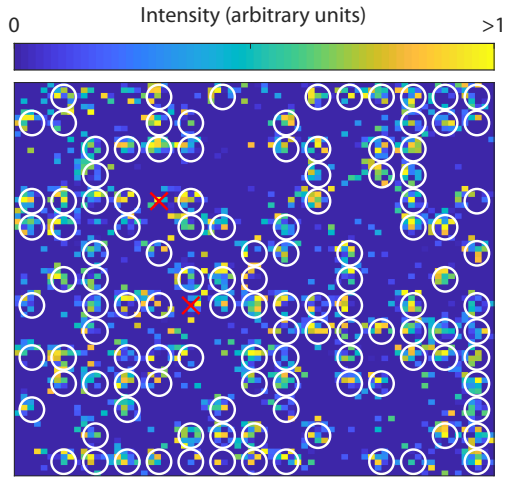}
    \caption{Identification of occupied sites in a simulated 15-by-15 lattice of Yb atoms, with pinning but without cooling, and 40\% filling of the lattice. Sites classified as occupied are identified by white circles, and incorrectly classified sites are marked with red crosses. Three of the 225 sites in this particular image were misclassified, corresponding to 98.8\% fidelity, consistent with the average fidelity achieved on a larger test set. While images were binarized prior to input to the network, setting the value of each nonzero pixel to 1, we display a normalized unbinarized image here.}
    \label{fig:kozuma_reconstruction}
\end{figure}

We subsequently seek to evaluate our reconstruction technique for the case of noncooled imaging in which atoms are nevertheless confined in a deep lattice during imaging. We use as our guideline the first known successful implementation of this scheme, performed by Miranda et al. \cite{miranda_site-resolved_2017} using Yb atoms in a lattice of period 543.5 nm. As in the case of our unconfined Er lattice, the Yb atoms will be heated during imaging, eventually displacing them from their original lattice site. This means that both systems require relatively short imaging pulses with high scattering rates. The addition of a pinning lattice, however, causes the atoms to remain confined in a smaller region, and for a longer period of time, before their eventual loss. The steep potential gradient at the nodes of the lattice also drives atoms away from these regions, reducing the photon density between sites compared to the unconfined case. In the experiment, Yb atoms are imaged on the $^1S_0 - ^1P_1$ transition at 399 nm during a 40 $\mu$s pulse, while confined in a lattice of depth 150 $\mu$K. With a scattering rate of $1.3 \times 10^7$ Hz, each atom scatters an average of 520 photons per imaging pulse \cite{miranda_site-resolved_2015}, of which 6.6\% are detected by the camera. The combined loss and hopping rate is 2.5\% per pulse. In our simulation we achieve a threshold-based fidelity of 98.6\% and a fidelity using the convolutional classifier of 98.8\%, representing a small but consistent reduction in the error rate. Figure \ref{fig:kozuma_reconstruction} shows an example of a simulated image of a 15-by-15 lattice, with occupied sites identified by a trained convolutional network and labelled. 

\section{\label{sec:conclusion}Conclusion}
The extension of site-resolved imaging of optical lattices to noncooled atoms will be a step forward in the flexibility of quantum gas microscopy. We have demonstrated the effectiveness of using both feedforward and convolutional neural networks for the analysis of noncooled lattice images, where low photon counts and atom movement limit the fidelity of traditional reconstruction techniques. We have shown that reconstruction is viable for completely unconfined erbium atoms, for which we can reduce the error rate by as much as half compared to state-of-the-art threshold reconstruction. We have also shown that the convolutional neural networks are able to perform consistently as well or better than threshold-based reconstruction for trapped atoms using the test case of pinned ytterbium atoms.
The neural networks designed for this task are flexible, and can be applied to any imaging system which can be sufficiently well-simulated to produce large labelled data sets to train the network. This reconstruction technique can be trivially extended to continuously cooled imaging systems, where it may prove advantageous in cases where atoms are separated by much less than the diffraction limit of the imaging system and only a small number of photons can be collected.

\ack
We wish to thank M. Greiner and the whole erbium team at Harvard for fruitful discussions of the challenge of erbium imaging, and M. Miranda for input relating to the simulation of noncooled ytterbium imaging. We also wish to thank the anonymous referees for their highly productive feedback which furthered the development of this work. The neural networks presented were trained using the HPC infrastructure LEO of the University of Innsbruck. This work is financially supported through an ERC Consolidator Grant (RARE, no. 681432) and a DFG/FWF (FOR 2247/PI2790), the SFB FoQuS (FWF Project No. F4016-N23), a NFRI Grant (MIRARE, No. \"OAW0600) from the Austrian Academy of Science and the Quantum Flagship PASQuanS (Grant no. 817482).

\appendix

\section{Imaging Simulation}

\begin{figure}[h]
    \centering
    \includegraphics[width=0.48\textwidth]{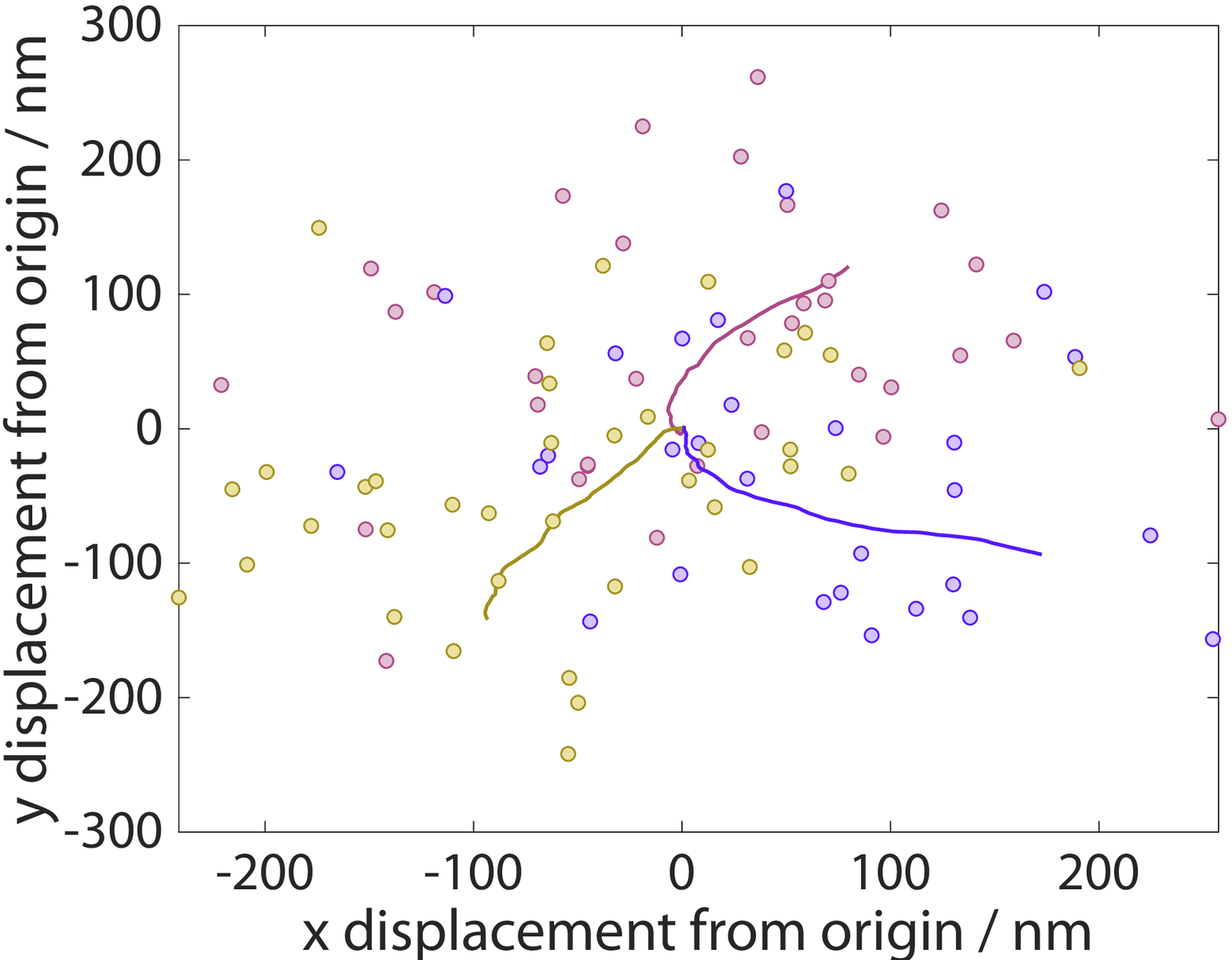}
    \caption{A set of three random walks for unconfined erbium atoms imaged with 401 nm light and illumination time 3 $\mu$s. The atom trajectories are marked by solid lines, and the photon detection positions are marked by circles, of the same color as their source atoms. Note that approximately six-times more photons are scattered than detected here.}
    \label{fig:rand_walk}
\end{figure}

Training our neural networks requires large labelled data sets of lattice images, which cannot feasibly be constructed using experimental data. As a result we use simulations of the imaging process in order to generate data sets of realistic images from arbitrary underlying lattice occupation patterns. The networks are trained on images of three-by-three site lattice segments, where only the central site is classified as occupied or unoccupied. Training data sets are made up of an equal number of simulated images of each of the 512 possible permutations of occupied and unoccupied sites in the three-by-three lattice. During classification of real images, the entire image will be divided up into overlapping three-by-three segments which are fed individually into the classifier network.

The simulation models the stochastic processes of photon scattering and atom movement which determine the image recorded by a quantum gas microscope. Atoms are assumed to begin at the center of each lattice site with zero velocity. We simulate scattering events in which photons are absorbed from four imaging beams aligned in the imaging plane and re-emitted in a random direction of the full solid angle ($4\pi$), creating a discrete velocity kick with the corresponding recoil momentum at each event. If there is a lattice potential switched on during imaging, the acceleration and velocity of the atom are updated according to the velocity Verlet algorithm. The lattice potential is assumed to be a symmetric $\sin^2$ potential with an amplitude (trap depth) given as an input parameter to the simulation. Emitted photons are recorded by the camera with a probability given by the collection efficiency, which is determined by the geometry of the imaging system, overall losses due to absorption and quantum efficiency of the camera. Each photon is detected at a random position around the location of the atom itself, with a probability distribution determined by a point spread function centered on the atom. In the initial phase of the simulation, each photon detection is represented by unity addition to the illuminated pixel in the simulated image.

The scattering code is looped with randomized exponential-distributed timesteps between absorption and re-emission, with the natural linewidth as input parameter, leading to an effective scattering rate at about half the natural linewidth as expected. The imaging process is concluded when the total elapsed time exceeds a given imaging pulse time or when the atom escapes the not fully closed transition cycle, accounted for by a small finite lossrate evaluated at each scattering event. Over the course of an imaging pulse, the accumulation of velocity kicks heats the atom and causes it to move on a random walk away from its initial position. Some example random walks are illustrated in figure \ref{fig:rand_walk}.

After looping over the imaging time for all atoms, Poissonian noise is added to each pixel of the image to account for clock-induced charges, with a mean noise value per pixel estimated from state-of-the-art EMCCD cameras. We also add an overlay of bright pixels consisting of the leak light from a random configuration of next-nearest neighbors to each image. Only 1000 such overlays are generated, as opposed to every five-by-five configuration, and they are randomly added to all images in the training data set. Finally, the electron multiplier gain from EMCCD cameras is applied to every pixel to calculate how many electrons per pixel will be present \cite{tubbs_lucky_2003}. The final conversion step into counts per pixel, requiring multiplication with a constant factor, adding a constant offset and including the electronic readout noise, was omitted in the present analysis. We also did not include further effects like additional charges due to background light or dark current as they should be negligible under the assumed experimental conditions.

Finally, we implement a preprocessing step, normalizing the data before feeding the data to the neural network for analysis. In the case of images with a low recorded photon count, where each pixel is very unlikely to be doubly illuminated, preprocessing consists of binarizing the images by setting the value of each illuminated pixel to 1 and all others to 0. For images with a higher photon count in which doubly illuminated pixels are likely to occur, pixels are normalized to the range \{0,1\} according to the formula
\begin{equation} 
\label{A1}
\mathbf{x}_\mathrm{norm} = \tanh\left(\tanh^{-1}(0.5)\frac{\mathbf{x}}{\mu_\mathrm{bright}}\right)
\end{equation}
where $\mathbf{x}$ is an image, or batch of images concatenated to form a single vector, and $\mu_\mathrm{bright}$ is the mean value of all the nonzero elements of $\mathbf{x}$.

\section{Optimizing network hyperparameters}

\begin{figure}
    \centering
    \includegraphics[width=0.48\textwidth]{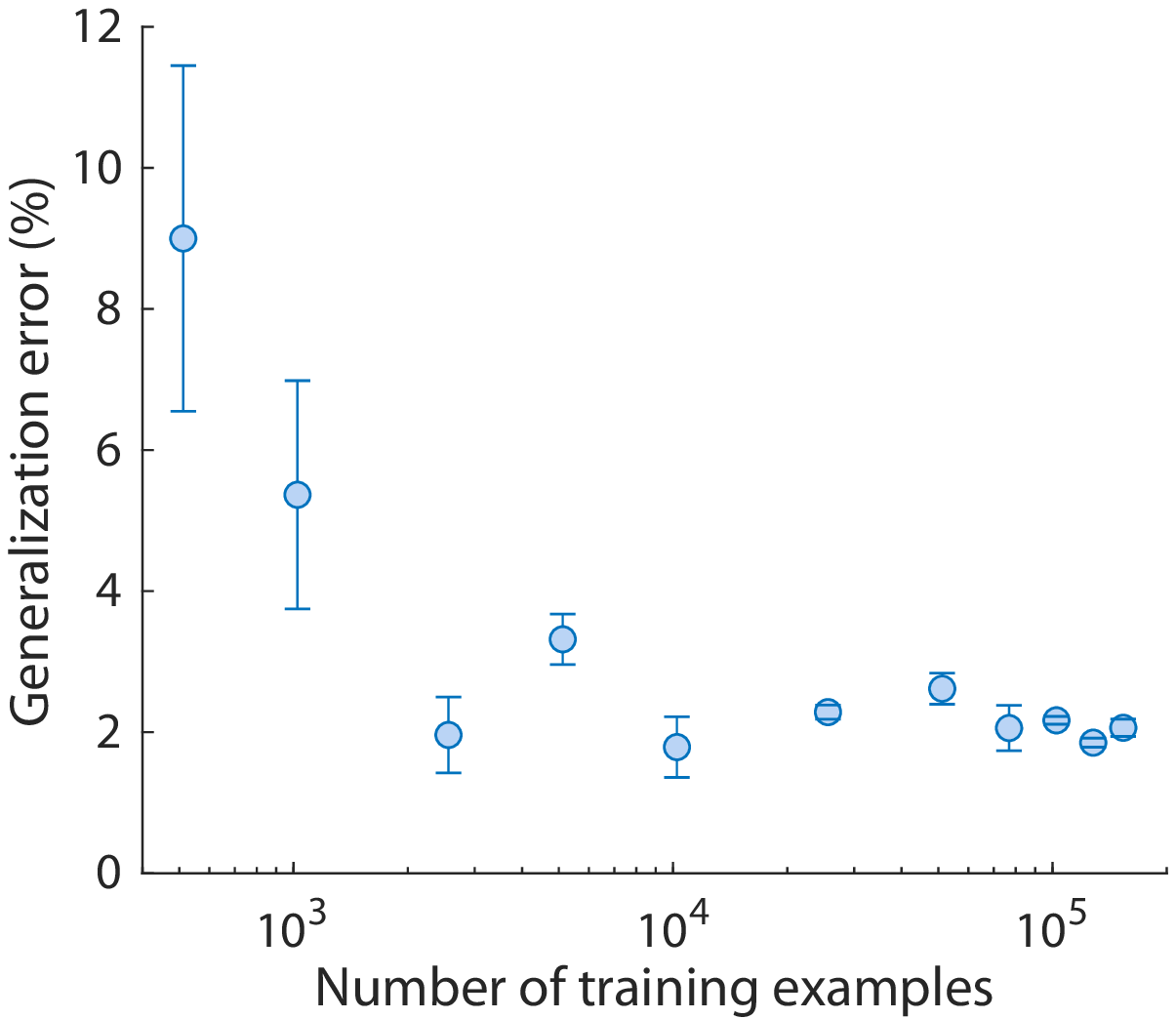}
    \caption{Convergence of convolutional neural network classification fidelity with increasing size of the training data set. All data sets are composed of a given number of repetitions of each possible occupation pattern of a three-by-three set of lattice sites, for erbium imaged at 401 nm for 3 $\mu$s.}
    \label{fig:NNConv}
\end{figure}

Hyperparameters are the parameters of the network which are not updated during training. Hyperparameters can be individually set by the architect of a neural network, or determined through a hyperoptimization process whereby multiple networks with different hyperparameters are separately trained and their performance compared to select the optimal hyperparameter values.

Aside from network architecture, the most significant hyperparameter in our case is the size of the training data set. We use data sets composed of equal numbers of simulated images generated from each of the 512 possible distributions of atoms in a three-by-three lattice segment. We trained both three-layer and convolutional networks on data sets consisting of between $10^3$ and $1.5\times10^5$ individual images of erbium lattice segments with 266 nm lattice spacing. As can be seen in figure \ref{fig:NNConv}, the generalization error of the convolutional network is minimized for $12.8\times10^5$ images, corresponding to 250 images for each possible distribution of atoms. The error of the three-layer network also generally decreased, though its error is less consistent between different datasets due to the difficulty of reliably converging to a good local minimum without prior dimensionality reduction. As the unconfined erbium atoms at 266 nm spacing represent the most difficult test case for our networks, it can be assumed that other cases would not need any larger training sets.

For the convolutional network, we also need to optimize a number of parameters for each convolutional and pooling layer. As described in the text, we find a kernel size of 10-by-10 pixels for all layers gives us our best performance. We use a progressively increasing number of filters in each convolutional level, beginning with 20 in the first layer followed by 40 in the second and 100 in the final layer. For the pooling layers, we find that our best performance is achieved for a pooling region of 5-by-5 pixels with a stride of 2.

\section*{References}
\bibliography{main.bib}

\begin{thebibliography}{10}

\bibitem{gross_quantum_2017}
Christian Gross and Immanuel Bloch.
\newblock Quantum simulations with ultracold atoms in optical lattices.
\newblock {\em Science}, 357(6355):995--1001, September 2017.

\bibitem{karski_nearest-neighbor_2009}
M.~Karski, L.~Förster, J.~M. Choi, W.~Alt, A.~Widera, and D.~Meschede.
\newblock Nearest-{Neighbor} {Detection} of {Atoms} in a 1d {Optical} {Lattice}
  by {Fluorescence} {Imaging}.
\newblock {\em Phys. Rev. Lett.}, 102(5):053001, February 2009.

\bibitem{sherson_single-atom-resolved_2010}
Jacob~F. Sherson, Christof Weitenberg, Manuel Endres, Marc Cheneau, Immanuel
  Bloch, and Stefan Kuhr.
\newblock Single-atom-resolved fluorescence imaging of an atomic {Mott}
  insulator.
\newblock {\em Nature}, 467(7311):68, September 2010.

\bibitem{bakr_quantum_2009}
Waseem~S. Bakr, Jonathon~I. Gillen, Amy Peng, Simon Fölling, and Markus
  Greiner.
\newblock A quantum gas microscope for detecting single atoms in a
  {Hubbard}-regime optical lattice.
\newblock {\em Nature}, 462(7269):74, November 2009.

\bibitem{cheuk_quantum-gas_2015}
Lawrence~W. Cheuk, Matthew~A. Nichols, Melih Okan, Thomas Gersdorf, Vinay~V.
  Ramasesh, Waseem~S. Bakr, Thomas Lompe, and Martin~W. Zwierlein.
\newblock Quantum-{Gas} {Microscope} for {Fermionic} {Atoms}.
\newblock {\em Phys. Rev. Lett.}, 114(19):193001, May 2015.

\bibitem{edge_imaging_2015}
G.~J.~A. Edge, R.~Anderson, D.~Jervis, D.~C. McKay, R.~Day, S.~Trotzky, and
  J.~H. Thywissen.
\newblock Imaging and addressing of individual fermionic atoms in an optical
  lattice.
\newblock {\em Phys. Rev. A}, 92(6):063406, December 2015.

\bibitem{haller_single-atom_2015}
Elmar Haller, James Hudson, Andrew Kelly, Dylan~A. Cotta, Bruno Peaudecerf,
  Graham~D. Bruce, and Stefan Kuhr.
\newblock Single-atom imaging of fermions in a quantum-gas microscope.
\newblock {\em Nat. Phys.}, 11(9):738, September 2015.

\bibitem{omran_microscopic_2015}
Ahmed Omran, Martin Boll, Timon~A. Hilker, Katharina Kleinlein, Guillaume
  Salomon, Immanuel Bloch, and Christian Gross.
\newblock Microscopic {Observation} of {Pauli} {Blocking} in {Degenerate}
  {Fermionic} {Lattice} {Gases}.
\newblock {\em Phys. Rev. Lett.}, 115(26):263001, December 2015.

\bibitem{parsons_site-resolved_2015}
Maxwell~F. Parsons, Florian Huber, Anton Mazurenko, Christie~S. Chiu, Widagdo
  Setiawan, Katherine Wooley-Brown, Sebastian Blatt, and Markus Greiner.
\newblock Site-{Resolved} {Imaging} of {Fermionic} $^6\mathrm{Li}$ in an
  {Optical} {Lattice}.
\newblock {\em Phys. Rev. Lett.}, 114(21):213002, May 2015.

\bibitem{mitra_quantum_2018}
Debayan Mitra, Peter~T. Brown, Elmer Guardado-Sanchez, Stanimir~S. Kondov,
  Trithep Devakul, David~A. Huse, Peter Schauß, and Waseem~S. Bakr.
\newblock Quantum gas microscopy of an attractive fermi–hubbard system.
\newblock {\em Nat. Phys.}, 14(2):173--177, February 2018.

\bibitem{nelson_imaging_2007}
Karl~D. Nelson, Xiao Li, and David~S. Weiss.
\newblock Imaging single atoms in a three-dimensional array.
\newblock {\em Nat. Phys.}, 3(8):556--560, August 2007.

\bibitem{alberti_super-resolution_2016}
Andrea Alberti, Carsten Robens, Wolfgang Alt, Stefan Brakhane, Michał Karski,
  René Reimann, Artur Widera, and Dieter Meschede.
\newblock Super-resolution microscopy of single atoms in optical lattices.
\newblock {\em New J. Phys.}, 18(5):053010, May 2016.

\bibitem{greif_site-resolved_2016}
Daniel Greif, Maxwell~F. Parsons, Anton Mazurenko, Christie~S. Chiu, Sebastian
  Blatt, Florian Huber, Geoffrey Ji, and Markus Greiner.
\newblock Site-resolved imaging of a fermionic {Mott} insulator.
\newblock {\em Science}, 351(6276):953--957, February 2016.

\bibitem{miranda_site-resolved_2017}
Martin Miranda, Ryotaro Inoue, Naoki Tambo, and Mikio Kozuma.
\newblock Site-resolved imaging of a bosonic {Mott} insulator using ytterbium
  atoms.
\newblock {\em Phys. Rev. A}, 96(4):043626, October 2017.

\bibitem{bergschneider_spin-resolved_2018}
Andrea Bergschneider, Vincent~M. Klinkhamer, Jan~Hendrik Becher, Ralf Klemt,
  Gerhard Zürn, Philipp~M. Preiss, and Selim Jochim.
\newblock Spin-resolved single-atom imaging of $^6${Li} in free space.
\newblock {\em Phys. Rev. A}, 97(6):063613, June 2018.

\bibitem{zdeborova_machine_2017}
Lenka Zdeborová.
\newblock Machine learning: {New} tool in the box.
\newblock {\em Nat. Phys.}, 13:420, February 2017.

\bibitem{mills_deep_2017}
Kyle Mills, Michael Spanner, and Isaac Tamblyn.
\newblock Deep learning and the {Schrödinger} equation.
\newblock {\em Phys. Rev. A}, 96(4):033402, October 2017.

\bibitem{seif_machine_2018}
Alireza Seif, Kevin~A. Landsman, Norbert~M. Linke, Caroline Figgatt, C.~Monroe,
  and Mohammad Hafezi.
\newblock Machine learning assisted readout of trapped-ion qubits.
\newblock {\em J. Phys. B: At. Mol. Opt. Phys.}, 51(17):174006, August 2018.

\bibitem{goy_low_2018}
Alexandre Goy, Kwabena Arthur, Shuai Li, and George Barbastathis.
\newblock Low {Photon} {Count} {Phase} {Retrieval} {Using} {Deep} {Learning}.
\newblock {\em Phys. Rev. Lett.}, 121(24):243902, December 2018.

\bibitem{chng_machine_2017}
Kelvin Ch’ng, Juan Carrasquilla, Roger~G. Melko, and Ehsan Khatami.
\newblock Machine {Learning} {Phases} of {Strongly} {Correlated} {Fermions}.
\newblock {\em Phys. Rev. X}, 7(3):031038, August 2017.

\bibitem{carrasquilla_machine_2017}
Juan Carrasquilla and Roger~G. Melko.
\newblock Machine learning phases of matter.
\newblock {\em Nat. Phys.}, 13(5):431, May 2017.

\bibitem{rem_identifying_2019}
Benno~S. Rem, Niklas Käming, Matthias Tarnowski, Luca Asteria, Nick
  Fläschner, Christoph Becker, Klaus Sengstock, and Christof Weitenberg.
\newblock Identifying quantum phase transitions using artificial neural
  networks on experimental data.
\newblock {\em Nat. Phys.}, 15:917, July 2019.

\bibitem{bohrdt_classifying_2019}
Annabelle Bohrdt, Christie~S. Chiu, Geoffrey Ji, Muqing Xu, Daniel Greif,
  Markus Greiner, Eugene Demler, Fabian Grusdt, and Michael Knap.
\newblock Classifying snapshots of the doped hubbard model with machine
  learning.
\newblock {\em Nat. Phys.}, 15:921, July 2019.

\bibitem{carleo_machine_2019}
Giuseppe Carleo, Ignacio Cirac, Kyle Cranmer, Laurent Daudet, Maria Schuld,
  Naftali Tishby, Leslie Vogt-Maranto, and Lenka Zdeborová.
\newblock Machine learning and the physical sciences.
\newblock {\em {arXiv}:1903.10563 [astro-ph, physics:cond-mat, physics:hep-th,
  physics:physics, physics:quant-ph]}, March 2019.

\bibitem{torlai_neural-network_2018}
Giacomo Torlai, Guglielmo Mazzola, Juan Carrasquilla, Matthias Troyer, Roger
  Melko, and Giuseppe Carleo.
\newblock Neural-network quantum state tomography.
\newblock {\em Nat. Phys.}, 14(5):447, May 2018.

\bibitem{ilzhofer_two-species_2018}
P.~Ilzhöfer, G.~Durastante, A.~Patscheider, A.~Trautmann, M.~J. Mark, and
  F.~Ferlaino.
\newblock Two-species five-beam magneto-optical trap for erbium and dysprosium.
\newblock {\em Phys. Rev. A}, 97(2):023633, February 2018.

\bibitem{goodfellow_deep_2016}
Ian Goodfellow, Yoshua Bengio, and Aaron Courville.
\newblock {\em Deep {Learning}}.
\newblock MIT Press, 2016.

\bibitem{bengio_towards_2015}
Yoshua Bengio, Dong-Hyun Lee, Jorg Bornschein, Thomas Mesnard, and Zhouhan Lin.
\newblock Towards {Biologically} {Plausible} {Deep} {Learning}.
\newblock {\em arXiv:1502.04156 [cs]}, February 2015.

\bibitem{lipton_mythos_2016}
Zachary~C. Lipton.
\newblock The mythos of model interpretability.
\newblock {\em {arXiv}:1606.03490 [cs, stat]}, June 2016.

\bibitem{matlab_deepdreamimage_2019}
MathWorks 2019.
\newblock {deepDreamImage. MATLAB R2019a documentation.
  [\textit{https://www.mathworks.com/\linebreak
  help/deeplearning/ref/deepdreamimage.html}] (Accessed August 2019)}.

\bibitem{aikawa_bose-einstein_2012}
K.~Aikawa, A.~Frisch, M.~Mark, S.~Baier, A.~Rietzler, R.~Grimm, and
  F.~Ferlaino.
\newblock Bose-{Einstein} {Condensation} of {Erbium}.
\newblock {\em Phys. Rev. Lett.}, 108(21):210401, May 2012.

\bibitem{aikawa_reaching_2014}
K.~Aikawa, A.~Frisch, M.~Mark, S.~Baier, R.~Grimm, and F.~Ferlaino.
\newblock Reaching {Fermi} {Degeneracy} via {Universal} {Dipolar} {Scattering}.
\newblock {\em Phys. Rev. Lett.}, 112(1):010404, January 2014.

\bibitem{miranda_site-resolved_2015}
Martin Miranda, Ryotaro Inoue, Yuki Okuyama, Akimasa Nakamoto, and Mikio
  Kozuma.
\newblock Site-resolved imaging of ytterbium atoms in a two-dimensional optical
  lattice.
\newblock {\em Phys. Rev. A}, 91(6):063414, June 2015.

\bibitem{tubbs_lucky_2003}
Robert~N. Tubbs.
\newblock {\em Lucky {Exposures}: {Diffraction} {Limited} {Astronomical}
  {Imaging} {Through} the {Atmosphere}}.
\newblock Doctoral, Cambridge University, Cambridge, September 2003.

\end{thebibliography}

\end{document}